\RequirePackage{ifpdf}
\documentclass{JINST} 
\usepackage{upgreek} \usepackage{enumitem} \usepackage{subfigure}\usepackage{amsmath}
\RequirePackage{lineno}

\title{Examining the Geometric Mean Method for the Extraction of Spatial Resolution}

\author{T. Alexopoulos$^a$, G. Iakovidis$^{a,b}$, S. Leontsinis$^{a,b}$
\thanks{Corresponding author.}
, K. Ntekas$^a$, V. Polychronakos$^b$\\
\llap{$^a$}National Technical University of Athens, Physics Department,\\
  9-Iroon Polytechniou, GR - 15780 Zografou, Greece\\
\llap{$^b$}Brookhaven National Laboratory, Physics Department,\\ Bldg. 510A, Upton, NY 11973, United States of America
\\E-mail: \email{Stefanos.Leontsinis@cern.ch}}

\abstract{
The spatial resolution of a detector, using a reference detector telecscope, can be measured applying the geometric mean method, with tracks reconstructed from hits of all the detectors, including ($\sigma_\mathrm{in}$) and excluding ($\sigma_\mathrm{ex}$) the hit from the detector under study. The geometric mean of the two measured resolution values ($\sigma=\sqrt{\sigma_\mathrm{ex}\sigma_\mathrm{in}}$), is proposed to provide a more accurate estimate of the intrinsic detector resolution. This method has been tested using a Monte Carlo algorithm and is proven to give accurate results, independently of the distance between the detectors used for the track fitting. The method does not give meaningful results if all the detectors do not carry the same characteristics.
}


\begin{document}

\section{Introduction}
The geometric mean method was proposed for calculating the spatial resolution of a GEM detector \cite{ref:theorypaper} and has been also adopted by performance studies of micromegas detectors \cite{ref_use1} \cite{ref_use2}. Apart from TPC like detectors this method can be applied in the case where the test detector is examined with the help of a set of precise detectors. This method is based on a straight track fit to the hits of all the reference telescope detectors including and excluding the hit of the detector under study (test detector). Calculating the standard deviation of the residuals, including the test detector in the fit, will bias the result in favour of smaller resolution values ($\sigma_{\mathrm{in}}$). Excluding the test detector hit from the fit will result in a systematically larger resolution ($\sigma_{\mathrm{ex}}$). Carnegie {\it et al.} \cite{ref:theorypaper} suggest that the true resolution, $\sigma$, is given by the geometrical mean of the two measurements:
\begin{equation}
\sigma^2 = \sigma_{\mathrm{in}} \sigma_{\mathrm{ex}}
\label{eq:formula}
\end{equation}
The proof of the formula can be found in appendix A of Reference \cite{ref:theorypaper} and appendix C of Reference \cite{ref_use1}.

A Monte Carlo (MC) algorithm is used to validate and test this method. It is shown that the geometrical mean combination of the two measurements can be used only when detectors with similar characteristics are included in the track fit.

\section{Monte Carlo Method}
\label{sec:mc}
The MC algorithm supposes four reference and one test detectors. The four reference detectors, are considered to have same spatial resolution and are positioned perpendicularly to the particle track separated by equal distances ($40\,\mathrm{cm}$). Three scenarios are studied where the resolution of the reference detectors is considered to be equal to $50\, \upmu\mathrm{m}$, $75\, \upmu\mathrm{m}$ and $100\, \upmu\mathrm{m}$, respectively.

\begin{figure}[h!]
	\begin{center}
		\includegraphics[width=0.6\textwidth]{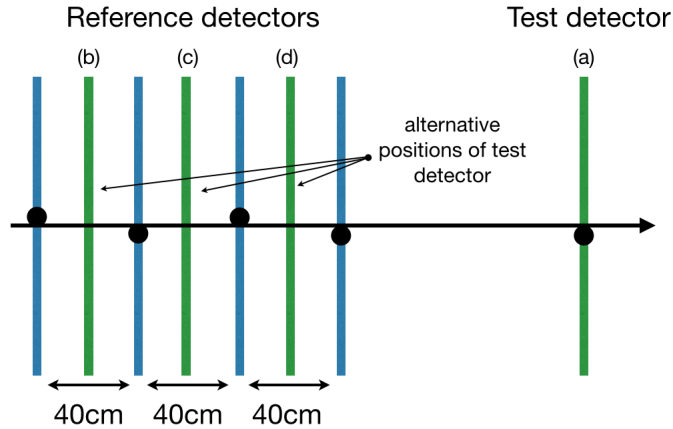}
		\caption{Configuration used for the modeling of the MC.}
		\label{fig:setup}
	\end{center}
\end{figure}

The test detector is positioned after the reference detectors setup, position (a) (see Figure \ref{fig:setup}). The geometric mean method is tested by varying the intrinsic resolution of the test detector from $35\,\upmu\mathrm{m}$ to $215\,\upmu\mathrm{m}$ in steps of $20\,\upmu\mathrm{m}$. The same exercise is repeated, changing the position of the test detector, (b), (c), (d), as shown in Figure \ref{fig:setup}.

In order to test the dependance of the geometric mean method as a function of the distance between the reference and the test detectors, one more scenario is considered. The intrinsic resolution of all five detectors is set to $70\,\upmu\mathrm{m}$. The distance of the test detector from the reference detectors is a variable fraction ($25\%$ to $475\%$ in steps of $25\%$) of the distance separating the reference detectors.

Single charged particles are simulated crossing the detectors creating single hits on each detector. These hits are smeared with a gaussian of a width equal to the detector resolution. Following the method described in Reference \cite{ref:theorypaper} we examine two track hypotheses. The first track is formed by fitting only the hits from the reference detectors and then a second track using all five hits coming from the reference and test detectors. By calculating the residuals of the test detector's hit position from the two formed tracks, the resolutions $\sigma_{\mathrm{ex}}$ and $\sigma_{\mathrm{in}}$ are extracted. Using the geometrical mean method (Equation \ref{eq:formula}) the spatial resolution of the test detector is calculated and compared to the generated one.

\begin{figure}[h]\begin{center}
		\includegraphics[width=0.45\textwidth]{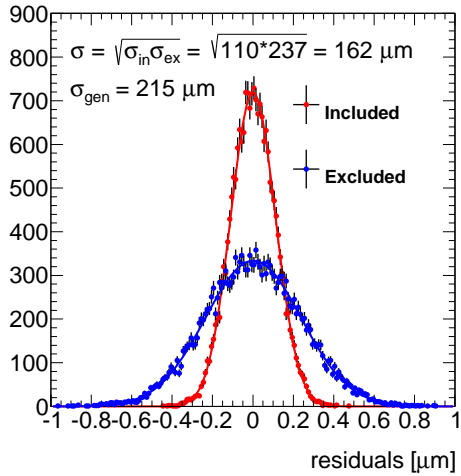}
		\caption{Spatial resolution plots using the two tracks, including (red points) and excluding (blue points) the hit from the test detector in the track fit. Both distributions are fitted with a gaussian function.}
		\label{fig:example_event}
\end{center}
\end{figure}

In Figure \ref{fig:example_event} the residual distributions for a sample of $10\, 000$ events is shown. The distribution with the blue points corresponds to the track fitted excluding the hit from the test detector and with the red points the distribution including it. From these distributions, the resolutions of the test detector are calculated to be $\sigma_{\mathrm{ex}}=237\, \upmu\mathrm{m}$ and $\sigma_{\mathrm{in}}=110\, \upmu\mathrm{m}$. The combination of these, results in an intrinsic resolution of $\sigma=162\, \upmu\mathrm{m}$, where the resolution used to generate the events was $215\, \upmu\mathrm{m}$ for the test detector and $100\, \upmu\mathrm{m}$ for the reference detectors.

\begin{figure}[h]\begin{center}
	\subfigure[] {
		\includegraphics[width=0.45\textwidth]{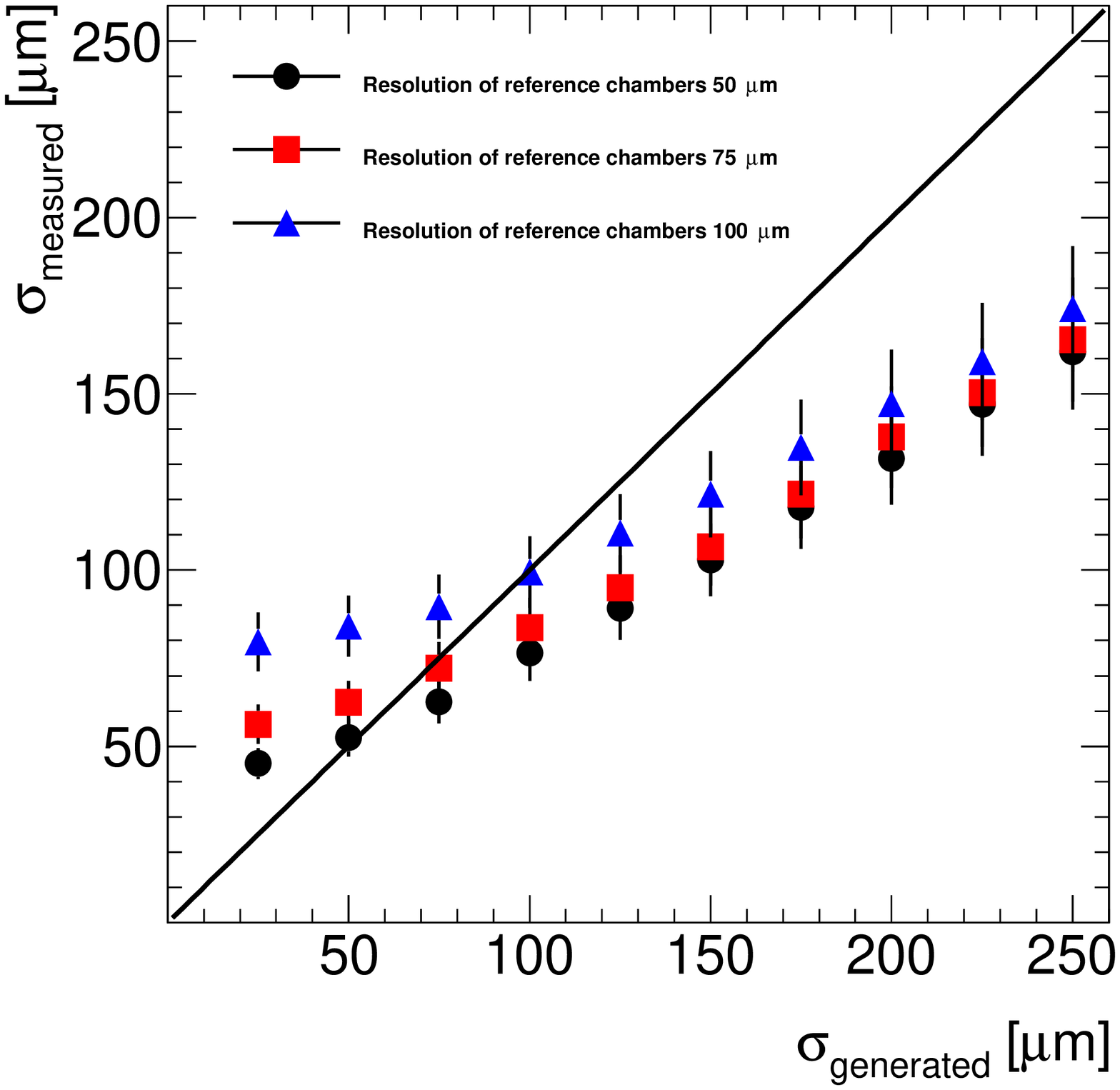}
		\label{fig:result}
	}
	\subfigure[] {
		\includegraphics[width=0.45\textwidth]{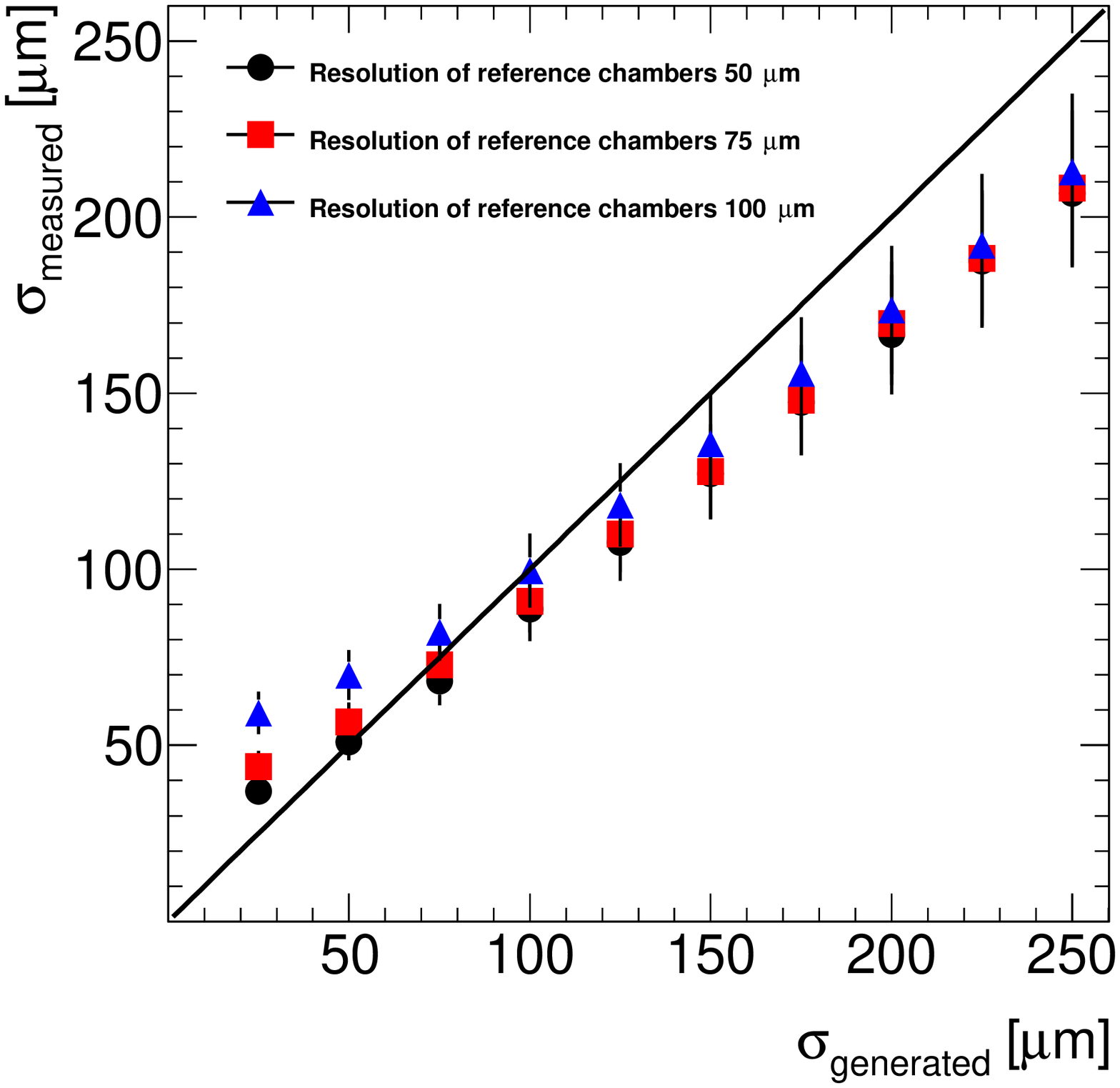}
		\label{fig:result_a}
	}
	\subfigure[] {
		\includegraphics[width=0.45\textwidth]{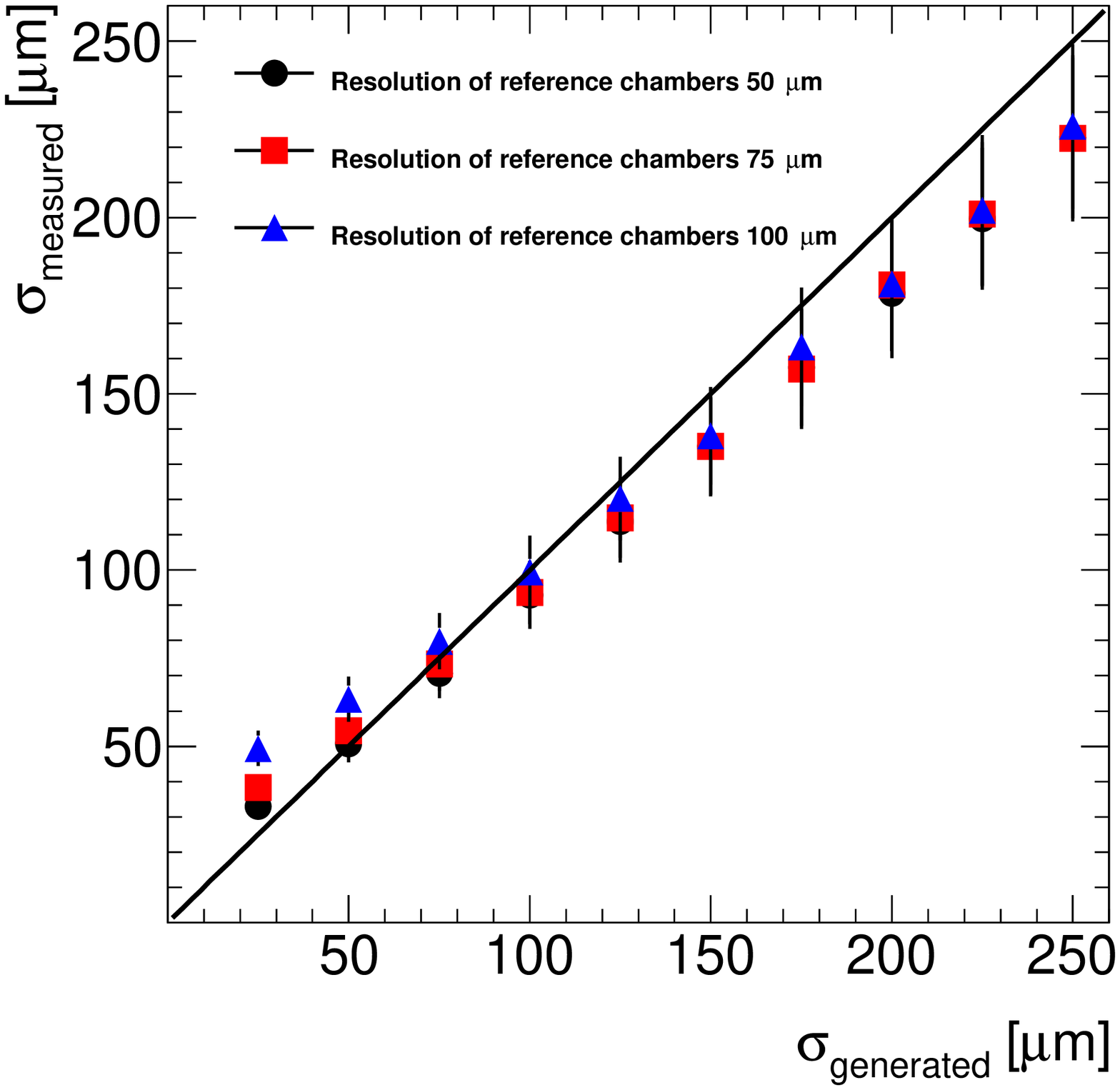}
		\label{fig:result_b}
	}
	\subfigure[] {
		\includegraphics[width=0.45\textwidth]{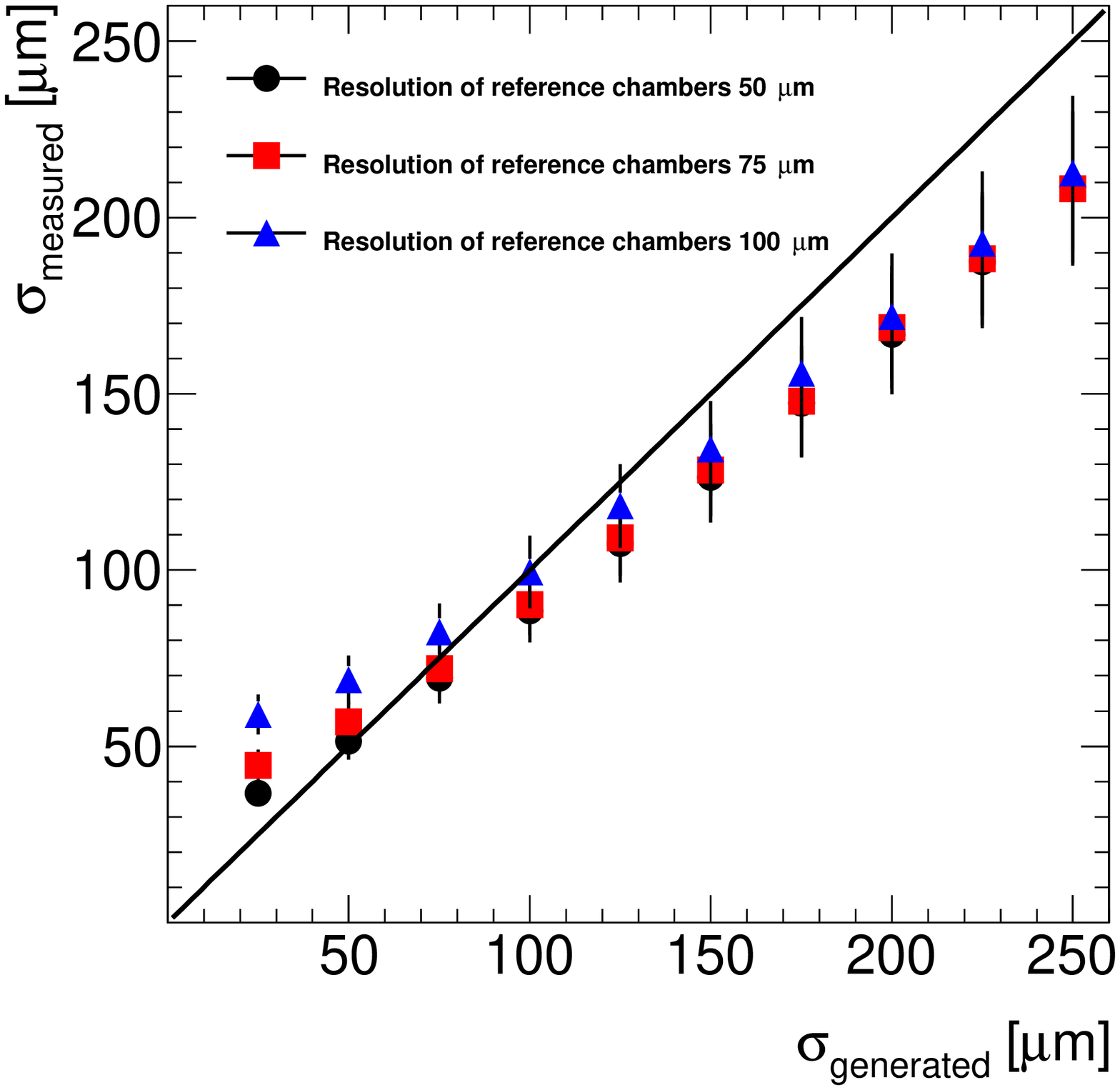}
		\label{fig:result_c}
	}
	\caption{(a) Comparison of measured and generated resolution, assuming three different scenarios. With the black circles the reference detectors are modelled to have $50\, \upmu\mathrm{m}$ resolution, with red squares $75\, \upmu\mathrm{m}$ resolution and with blue triangles $100\, \upmu\mathrm{m}$ resolution. The black line assumes that the generated value is equal to the measured. (b), (c), (d) Same as Figure \protect\ref{fig:result}, but the test chamber positioned in the alternative positions (b), (c) and (d), respectively (see Figure \protect\ref{fig:setup}). The errors on all figures are multiplied by 10 in order to be visible.}
	\label{fig:both_results}\end{center}
\end{figure}

\section{Results}
The results for three different values ($50\, \upmu\mathrm{m}$, $75\, \upmu\mathrm{m}$ and $100\, \upmu\mathrm{m}$) of generated resolution are summarised in Figure \ref{fig:result}. Each of the three scenarios is shown with different colour points. On the $x$ axis the generated resolution of the test detector is shown and the $y$ axis corresponds to the measured resolution using the geometric mean method. The diagonal black line indicates the occasion where the generated resolution is equal to the calculated. It is clearly shown that the results / points cross the line only when the generated resolution of the test detector matches the resolution of the reference detectors. When the true resolution of the test detector is worse than the references' the result is biased showing better resolution than in reality.

The same behaviour is observed when the test detector is placed in the alternative positions (b), (c) and (d), respectively (see Figure \ref{fig:setup}). The result of each position can be seen in Figures \ref{fig:result_a}, \ref{fig:result_b} and \ref{fig:result_c}. Occasions (b) and (d) place the test detector in two symmetric configurations producing the same results, as expected (Figures \ref{fig:result_a} and \ref{fig:result_c}).

\begin{figure}[h]\begin{center}
		\includegraphics[width=0.45\textwidth]{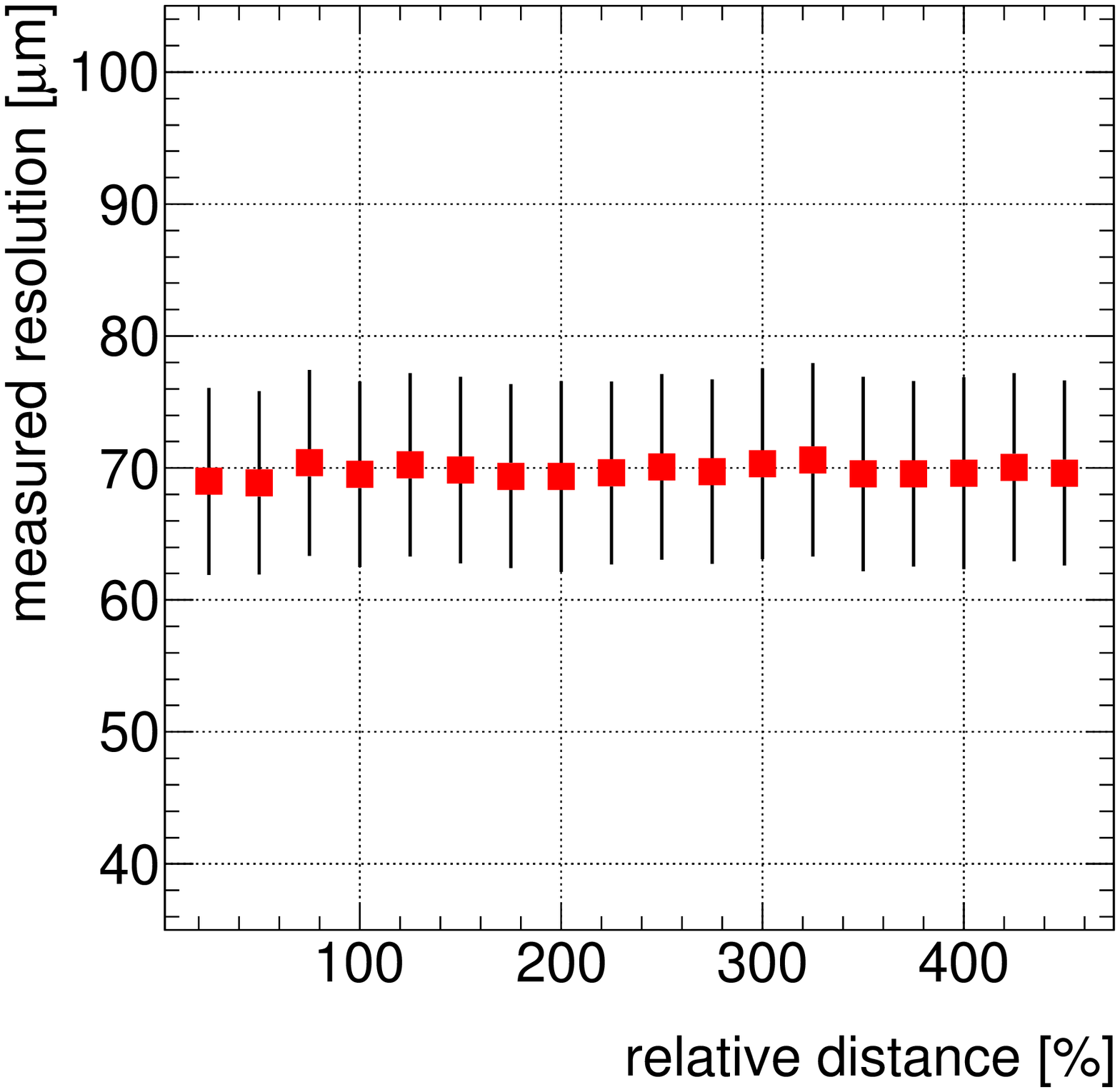}
		\caption{Calculated resolution using the geometric mean method as a function of the distance of the test detector from the reference detectors. Errors on both figures are multiplied by 10 in order to be visible.}
		\label{fig:distance}
\end{center}
\end{figure}

The distance between the test and reference detectors does not affect the measurement of the resolution. Figure \ref{fig:distance} shows that the measured spatial resolution of the test detector remains constant with respect to its distance from the reference setup.

\section{Conclusions}
\label{sec:conclusions}
The geometric mean method produces accurate results when the test and reference detectors have the same characteristics. However, when the resolution of the test detector is worse than the reference ones, the result is biased towards better performance. This behaviour is observed in both cases where the test detector is placed inside the reference detectors setup and in the outside area. Finally, it is shown that the distance between the test and reference detectors does not affect the calculated spatial resolution.

\section*{Acknowledgments}
The present work was co-funded by the European Union (European Social Fund ESF) and Greek national funds through the Operational Program "Education and Lifelong Learning" of the National Strategic Reference Framework (NSRF) 2007-1013. ARISTEIA-1893-ATLAS MICROMEGAS.

\end{document}